\documentclass[11pt]{article}
\usepackage[margin=1in]{geometry}
\usepackage{graphicx}
\usepackage{overpic}
\usepackage{amsmath,amssymb,amsfonts}
\usepackage{amsthm}
\usepackage{mathrsfs}
\usepackage{cite}
\usepackage{booktabs}
\usepackage{multirow}
\usepackage{float}
\usepackage{algorithm}
\usepackage{algpseudocode}
\usepackage{listings}
\usepackage{xcolor}
\usepackage{textcomp}
\usepackage{placeins}
\usepackage{needspace}
\usepackage{subfig}
\usepackage{authblk}
\usepackage{hyperref}
\usepackage{pict2e}

\newcommand{\Rb}{\mathbb{R}}
\newcommand{\xvec}{\mathbf{x}}
\newcommand{\yvec}{\mathbf{y}}
\newcommand{\zvec}{\mathbf{z}}
\newcommand{\svec}{\mathbf{s}}
\newcommand{\Enc}{\mathcal{E}}
\newcommand{\Dec}{\mathcal{D}}

\title{Replicating chaotic dynamics from sparse sensors with recurrent latent states}
\author[1]{Jan P. Williams}
\author[2]{J. Nathan Kutz}
\author[1]{Krithika Manohar}
\affil[1]{Department of Mechanical Engineering, University of Washington, Seattle, WA USA}
\affil[2]{Autodesk Research, London, UK}
\date{}

\begin{document}

\maketitle

%=====================================================================
\begin{abstract}
\noindent
Forecasting a high-dimensional nonlinear system from a few point sensors requires an estimate of the current state and a model of its evolution. Reduced-order models usually take the full state as input, so with only sparse sensors they require an auxiliary full-state estimate whose error is inherited by the latent representation. We instead build the latent state from sensor measurements alone, using a shallow recurrent decoder (SHRED) network that encodes a history of point measurements, and train an echo state network to advance that latent state autonomously. We show that the resulting latent dynamics replicate the long-term dynamical invariants of the underlying system rather than merely tracking its short-term trajectory. For the Kuramoto--Sivashinsky equation, the autonomous latent model reproduces the leading Lyapunov exponents of the full system from $25$ of $128$ state components, using a latent state with $6$--$14\%$ as many components as the full state. Because Lyapunov exponents are invariant under smooth coordinate changes, this is consistent with the recurrent encoder recovering an embedding of the attractor, in the spirit of Takens' delay embedding theorem. The recurrent encoding also decouples forecast skill from reconstruction accuracy. SHRED reconstructs less accurately than a convolutional autoencoder given the full state, yet it forecasts comparably on the simpler chaotic case and better on the more complex one. For Kolmogorov flow, with sensors at $6\%$ of grid points and a latent state $0.3\%$ the size of the vorticity field, the autonomous model remains on the learned latent manifold over long horizons.
\end{abstract}

\noindent\textbf{Keywords:} sparse sensing $\cdot$ reduced-order
modeling $\cdot$ echo state networks $\cdot$ Lyapunov spectrum $\cdot$
delay embedding $\cdot$ spatiotemporal chaos

%=====================================================================
\section{Introduction}
\label{sec:intro}

The prediction of spatiotemporally chaotic systems from limited measurements is a recurring problem throughout the engineering sciences. It arises in turbulence and combustion, in structural health monitoring, and in geophysical forecasting. In almost every practical setting, full state measurements are infeasible and sparse measurements must be used to infer fields with state dimension from $10^4$ up to $10^{11}$. Forecasting therefore cannot be separated from estimation, and at deployment, one usually has neither a governing equation nor a snapshot of the full state.

These two halves of the problem have mostly been studied on their own. On the estimation side, many standard approaches reconstruct the field at one instant from measurements at that same instant, either by projecting onto a precomputed basis, as in gappy proper orthogonal decomposition \cite{everson_karhunenloeve_1995} and empirical interpolation \cite{barrault_empirical_2004, chaturantabut_nonlinear_2010, drmac_new_2016} together with the sensor placement strategies built on them \cite{manohar_data-driven_2018, manohar_optimized_2019, clark_greedy_2019, desilva_pysensors_2021}, or by nonlinear estimators such as shallow decoder networks \cite{erichson_shallow_2020} and sparse representation in a library of training snapshots \cite{callaham_robust_2019}. In either case the reconstruction is instantaneous and memoryless. On the modeling side, methods learn evolution operators from data. These include dynamic mode decomposition and its Koopman-theoretic extensions \cite{schmid_dynamic_2010, williams_datadriven_2015, kutz_dynamic_2016}, sparse regression for governing equations \cite{brunton_discovering_2016, kaiser_sparse_2018}, and deep autoencoders that find coordinates in which the dynamics admit a parsimonious~\cite{kutz2022parsimony} representation \cite{hinton_reducing_2006, lusch_deep_2018, lee_model_2020, doi:10.1073/pnas.1906995116, linot_deep_2020, linot_datadriven_2022, gin2021deep}. Neural operators go further and learn maps between function spaces, such as the operator taking an initial condition or forcing to the solution of a partial differential equation at a later time \cite{chen_universal_1995, lu_learning_2021, li_fourier_2021, kovachki_neural_2023}. All of these methods assume access to the state they are asked to advance, or to a well-resolved sample of the input function.

Composing an estimation model and a dynamics model yields a reduced order model (ROM) that estimates the field from sensors, encodes that reconstruction, and then steps forward in the latent space. However, there is a structural problem with this approach. An autoencoder's encoder is a function of the instantaneous full state, so a sensor-limited deployment has to solve an ill-posed inverse problem at every step just to give the encoder an input. The errors that inversion introduces enter the latent state and are then amplified by the same chaotic dynamics the model is supposed to capture. The problem is exacerbated as the ratio of state dimension to sensor count grows. That ratio is large in most cases where one would want to use these methods at all. {\"O}zalp et al.\ \cite{OZALP2026118600} address this by wrapping the latent model in a sequential estimator, pairing a convolutional autoencoder and an echo state network \cite{racca_predicting_2023} with an ensemble Kalman filter that updates the latent state from sparse noisy measurements. They obtain stable real-time forecasts of the Kuramoto--Sivashinsky equation and of Kolmogorov flow. While the filter works, it adds complexity to compensate for an architecture that cannot take measurements directly.

Another option is to build the latent representation directly from measurements. The shallow recurrent decoder (SHRED) architecture \cite{williams_sensing_2024, ebers2023leveraging} does exactly this, reading a time-lagged trajectory from a few point sensors and producing a latent state from which a shallow decoder reconstructs the full field. Because the encoder takes a trajectory rather than a snapshot, it never needs the full state; its only inputs are measurements that are available at deployment. SHRED has been used for sensing and reconstruction in fluid, geophysical and biomechanical systems \cite{williams_sensing_2024, ebers2023leveraging}, extended to parametric reduced-order modeling \cite{tomasetto_reduced_2025}, and combined with sparse regression to produce interpretable latent models \cite{gao_sindy2026}. In most of this work SHRED serves as a reconstruction tool, and its latent state is an intermediate step toward the field at the current time. Here we ask a different question, namely whether the latent trajectory itself carries the dynamics of the underlying system.

There also exists theoretical justification for the use of a recurrent encoder. Takens' embedding theorem \cite{takens_detecting_1981} guarantees that for a generic scalar observable, the map from the state of a dynamical system to a long enough delay vector is an embedding, that is, a diffeomorphism onto its image. The result extends to fractal attractors \cite{sauer_embedology_1991} and to delay vectors built from several observables \cite{deyle_generalized_2011}. A delay-coordinate reconstruction therefore preserves the attractor up to a smooth change of coordinates. A recurrent network fed a lagged sensor trajectory is carrying out a learned, nonlinear version of the same construction. Uribarri and Mindlin \cite{uribarri_dynamical_2022} showed empirically that recurrent networks trained to forecast a single observable develop internal states topologically equivalent to the original strange attractor, and Bakarji et al.\ \cite{bakarji2022discovering} learned coordinates and governing equations from delay-embedded partial measurements with deep autoencoders. This matters because the invariants of a dynamical system, such as Lyapunov exponents, are preserved under smooth invertible changes of coordinate. Consequently, a latent representation learned from sparse sensors can in principle provide a diffeomorphic image of the attractor, and a time stepper trained on that representation can then reproduce the Lyapunov spectrum of the original system. Testing this possibility is the main purpose of this paper.

Echo state networks \cite{jaeger_echo_2001, maass_real-time_2002, lukosevicius2009reservoir} are a natural architecture for propagating the learned latent state of a SHRED model. An ESN drives a fixed random recurrent reservoir with the input signal and fits only a linear readout. Training therefore reduces to a linear least-squares problem, and the resulting map has a closed-form Jacobian, which is needed to compute Lyapunov exponents. ESNs also forecast chaotic systems well \cite{pathak_model-free_2018, vlachas_backpropagation_2020}, and Pathak et al.\ \cite{pathak_using_2017} showed the stronger result that a reservoir trained on the Kuramoto--Sivashinsky equation reproduces its Lyapunov spectrum. Their reservoir was driven by the full state, however. {\"O}zalp et al.\ \cite{ozalp2023reconstruction} showed that long short-term memory networks trained on partial observations of the Kuramoto--Sivashinsky equation can also recover its Lyapunov exponents, using either full-state targets in the loss or the governing equations. Whether spectrum replication holds when an autonomous time stepper sees only a low-dimensional latent state distilled from a handful of point sensors has not, to our knowledge, been established.

We show that it does, and that the latent coordinates produced by a recurrent encoder are better suited to forward propagation than those of a more accurate static autoencoder. The contributions of this paper are as follows.
\begin{itemize}

  \item \textbf{Dynamics replication from sparse observations.} An autonomous model trained on the latent states of a SHRED network reproduces the leading Lyapunov exponents of the Kuramoto--Sivashinsky equation, although its encoder receives only $25$ of the $128$ state components as input. Because the Lyapunov spectrum is invariant under smooth changes of coordinates and is not implied by short-term forecast accuracy, this agreement indicates that the latent coordinates capture the dynamical structure of the attractor.

  \item \textbf{Reconstruction accuracy does not determine forecastability.} At equal latent dimension, SHRED reconstructs the state less accurately than a convolutional autoencoder (CAE) given full-state access, though far more accurately than a linear POD projection. The SHRED-based forecasting pipeline nevertheless performs comparably to the CAE-based one on the simpler chaotic regime and outperforms it on the more complex one. We conjecture that the recurrent encoder produces temporally coherent latent trajectories that are easier to propagate, as argued in Section~\ref{sec:forecasting} (Figs.~\ref{fig:forecast_22} and~\ref{fig:forecast_44}).

  \item \textbf{Extension to a severely undersampled two-dimensional flow.} For Kolmogorov flow observed at about $6\%$ of grid points, the autonomous latent model remains on the invariant manifold traced by the encoded trajectory over long horizons, as measured by $\ell_2$ norm, neither blowing up nor collapsing to a fixed point.

\end{itemize}

The rest of the paper is organized as follows. Section~\ref{sec:methods} describes the architecture, the latent time stepper, and how we compute Lyapunov spectra. Section~\ref{sec:systems} introduces the two benchmark systems and the data generation protocol. Section~\ref{sec:results} summarizes reconstruction accuracy and then presents forecasting, spectrum replication and latent manifold results. Section~\ref{sec:discussion} discusses implications and future work, and Appendix~\ref{app:recon} gives the full reconstruction comparison as a function of latent dimension.

%=====================================================================
\section{Methods}
\label{sec:methods}

\subsection{Problem statement and notation}
\label{sec:problem}

Consider a dynamical system whose spatially discretized state at time $t_k$ is $\xvec_k \in \Rb^{n}$, evolving under an unknown flow map $\xvec_{k+1} = \Phi(\xvec_k)$. We observe the system only through $p \ll n$ fixed point sensors,
\begin{equation}
  \yvec_k = C\,\xvec_k \in \Rb^{p},
  \label{eq:measurement}
\end{equation}
where $C \in \Rb^{p \times n}$ selects $p$ components of the state vector at randomly chosen spatial locations that are held fixed for all time. Given a window of past measurements $\yvec_{k-L_{\mathrm{lag}}+1}, \ldots, \yvec_{k}$ and no knowledge of $\Phi$, we want to forecast $\xvec_{k+j}$ for $j = 1, 2, \ldots$

We accomplish this task through the use of a latent state $\zvec_k \in \Rb^{r}$ with $r \ll n$. An encoder $\Enc$ maps the measurement window to the latent state, a decoder $\Dec$ maps the latent state to the full field, and a latent flow map $F$ advances the latent state:
\begin{equation}
  \zvec_k = \Enc\!\left(\yvec_{k-L_{\mathrm{lag}}+1{:}k}\right),
  \qquad
  \hat{\xvec}_k = \Dec(\zvec_k),
  \qquad
  \zvec_{k+1} = F(\zvec_k).
  \label{eq:pipeline}
\end{equation}
The three maps are learned in two stages. First $\Enc$ and $\Dec$ are trained jointly by supervised reconstruction, then $F$ is fit to the latent trajectory $\Enc$ produces. Training requires full-state snapshots as targets in the reconstruction loss, as it does for an autoencoder, but $\Enc$ never takes $\xvec$ as an argument. This is what separates the approach from an autoencoder-based reduced-order model, whose encoder is a function of $\xvec_k$ alone, and it is the source of the results reported below.

\subsection{Shallow recurrent decoder network}
\label{sec:shred}

For the encoder and decoder we use a shallow recurrent decoder (SHRED) network \cite{williams_sensing_2024}. The encoder is a gated recurrent unit (GRU) \cite{cho_learning_2014, chung_empirical_2014} that reads the measurement window one step at a time. Writing $\mathbf{h}_j \in \Rb^{d_h}$ for the hidden state after ingesting $\yvec_j$, the GRU update is
\begin{align}
  \mathbf{u}_j &= \sigma\!\left(W_u \yvec_j + U_u \mathbf{h}_{j-1} + \mathbf{b}_u\right), \\
  \mathbf{q}_j &= \sigma\!\left(W_q \yvec_j + U_q \mathbf{h}_{j-1} + \mathbf{b}_q\right), \\
  \tilde{\mathbf{h}}_j &= \tanh\!\left(W_h \yvec_j + U_h (\mathbf{q}_j \odot \mathbf{h}_{j-1}) + \mathbf{b}_h\right), \\
  \mathbf{h}_j &= (1 - \mathbf{u}_j) \odot \mathbf{h}_{j-1} + \mathbf{u}_j \odot \tilde{\mathbf{h}}_j,
  \label{eq:gru}
\end{align}
where $\sigma$ is the logistic sigmoid, $\odot$ denotes elementwise multiplication, and $\mathbf{u}_j$ and $\mathbf{q}_j$ are the update and reset gates. We take the latent state of dimension $r$ to be $\zvec_k = W_{lat} \mathbf{h}_k$ where $W_{lat}$ is a learnable matrix. We found that this approach allowed for smaller $r$ than merely taking the final hidden state of the GRU. The gating in \eqref{eq:gru} mitigates the vanishing gradient problem that afflicts vanilla recurrent networks \cite{bengio_learning_1994, pascanu_difficulty_2013} and lets the latent state retain information across the whole lag window.

Two features of this construction are worth highlighting. First, $\zvec_k$ is a nonlinear function of the delay vector $(\yvec_{k-L_{\mathrm{lag}}+1}, \ldots, \yvec_k)$, and the delay vector is precisely the object to which Takens' theorem \cite{takens_detecting_1981} applies: for generic observables and $L_{\mathrm{lag}}$ sufficiently large, the delay map is an embedding of the attractor. Second, consecutive latent states are close by construction. In training, consecutive latent states are computed from windows that share $L_{\mathrm{lag}}-1$ measurements, and $\Enc$ is a continuous function of its window, so $\zvec_k$ and $\zvec_{k+1}$ are close. At deployment, \eqref{eq:gru} runs continuously as a discrete dynamical system on $\Rb^{d_h}$ driven by the sensor stream (Section~\ref{sec:esn}). In both cases the sequence $\zvec_1, \zvec_2, \ldots$ is a gradually varying function of the measurement history rather than a series of independently computed codes. In contrast, the latent code $\zvec_k$ in a convolutional autoencoder is computed from $\xvec_k$ alone, with no dependence on $\zvec_{k-1}$.

The decoder $\Dec$ maps $\zvec_k$ back to the full field. Following \cite{gao_sindy2026} we use transposed convolutional layers rather than the fully connected layers of the original SHRED formulation, which exploits the spatial structure of the fields studied here and cuts the parameter count. We train the network end to end by minimizing the mean squared reconstruction error
\begin{equation}
  \mathcal{L} = \frac{1}{m}\sum_{k=1}^{m}
  \left\| \Dec\!\left(\Enc\!\left(\yvec_{k-L_{\mathrm{lag}}+1{:}k}\right)\right) - \xvec_k \right\|_2^2
  \label{eq:shredloss}
\end{equation}
over the training snapshots, using Adam \cite{KingmaBa2014}. Note that this loss constrains the reconstruction and nothing else. No term encourages the latent trajectory to be smooth or low-order or otherwise well behaved. Any dynamical structure the latent space turns out to have comes from the recurrent parameterization rather than from an explicit penalty.  This makes the comparison in Section~\ref{sec:forecasting} informative; both architectures are trained against the same objective and differ only in how the latent state is built.

\subsection{Latent time stepper}
\label{sec:esn}

Once the SHRED model is trained we run the encoder over the full training set to extract the latent sequence $\zvec_1, \ldots, \zvec_m$ and standardize it to zero mean and unit variance coordinatewise. We then train an echo state network \cite{jaeger_echo_2001, lukosevicius_practical_2012} on this sequence to realize the autonomous map $F$ in \eqref{eq:pipeline}. The reservoir state $\svec_k \in \Rb^{N}$ evolves as
\begin{equation}
  \svec_{k+1} = (1-\alpha)\,\svec_k
  + \alpha \tanh\!\left(A\,\svec_k + \sigma_{\mathrm{in}} W_{\mathrm{in}} \zvec_k + \mathbf{b}\right),
  \label{eq:esnstate}
\end{equation}
where $A \in \Rb^{N \times N}$ is a sparse random adjacency matrix rescaled to spectral radius $\rho$, $W_{\mathrm{in}}$ is a random input matrix, $\sigma_{\mathrm{in}}$ is the input scaling, and $\alpha \in (0,1]$ is the leaking rate. The matrices $A$, $W_{\mathrm{in}}$ and the bias $\mathbf{b}$ are fixed at initialization and never trained. Only the linear readout is fit,
\begin{equation}
  \zvec_{k+1} = W_{\mathrm{out}} \svec_{k+1},
  \label{eq:esnreadout}
\end{equation}
by ridge regression with Tikhonov parameter $\beta$, giving the closed-form solution $W_{\mathrm{out}} = Z S^{\top}(S S^{\top} + \beta I)^{-1}$, where $S$ and $Z$ collect the reservoir states and the target latent states. We discard a spinup of $N_{\mathrm{spin}}$ steps before the regression so that $\svec$ is independent of its initialization, as the echo state property guarantees. In autonomous operation the readout \eqref{eq:esnreadout} is fed back into \eqref{eq:esnstate} in place of the true $\zvec_k$, which closes the loop. The resulting autonomous system is a map on the reservoir state,
\begin{equation}
  \svec_{k+1} = G(\svec_k) = (1-\alpha)\,\svec_k + \alpha\tanh\!\Big(\big(A + \sigma_{\mathrm{in}}W_{\mathrm{in}}W_{\mathrm{out}}\big)\svec_k + \mathbf{b}\Big),
  \label{eq:Gmap}
\end{equation}
from which the latent forecast is read out as $\zvec_k = W_{\mathrm{out}}\svec_k$. The latent flow map $F$ in \eqref{eq:pipeline} is therefore realized through the reservoir, with $W_{\mathrm{out}}\,G(\svec_k)$ advancing $\zvec_k = W_{\mathrm{out}}\svec_k$ to $\zvec_{k+1}$. We select the hyperparameters $(N, \rho, \sigma_{\mathrm{in}}, \alpha, \beta)$ by grid search on a validation set held out from the training data, using the GPU-accelerated implementation of \cite{williams_orc_2026}. We chose an ESN over alternative approaches because $G$ in \eqref{eq:Gmap} is an explicit composition of a fixed linear map and an elementwise nonlinearity, so its Jacobian is available in closed form. That is what makes the Lyapunov computation of Section~\ref{sec:lyap} tractable.

At deployment the pipeline runs as follows. Rather than re-encoding a fixed window at every step, the GRU runs continuously on the incoming sensor stream, carrying its hidden state forward, so the history it has seen grows with time. Because $L_{\mathrm{lag}}$ is taken long enough in training, the encoder trained on fixed-length windows can be used in this streaming mode. Its latent outputs warm up the ESN, which then iterates \eqref{eq:Gmap} forward on its own, and the decoder recovers $\hat{\xvec}_k = \Dec(W_{\mathrm{out}}\svec_k)$ at any point along the resulting trajectory. No full-state information enters at any stage of deployment.

\subsection{Lyapunov spectrum of the learned latent dynamics}
\label{sec:lyap}

Our main diagnostic is the Lyapunov spectrum, which we use to ask whether the learned latent dynamics are dynamically equivalent to the true dynamics rather than just predictive over short horizons.

Let $F$ be a smooth map on $\Rb^{d}$ that advances the state by one sampling interval $\Delta t$, with a trajectory $\{\zvec_k\}$ on an attractor, and let $J_T = \left.\partial F/\partial \zvec\right|_{\zvec_{T-1}} \cdots \left.\partial F/\partial \zvec\right|_{\zvec_{0}}$ be the product of Jacobians along that trajectory. Oseledets' multiplicative ergodic theorem \cite{eckmann_ergodic_1985} guarantees that $\Lambda = \lim_{T\to\infty} (J_T^{\top} J_T)^{1/(2T)}$ exists for almost every initial condition with respect to the invariant measure, and the Lyapunov exponents are
\begin{equation}
  \lambda_i = \frac{1}{\Delta t}\ln \mu_i(\Lambda) = \lim_{T \to \infty} \frac{1}{T\,\Delta t} \ln \sigma_i(J_T),
  \label{eq:lyapdef}
\end{equation}
where $\mu_i(\Lambda)$ is the $i$th eigenvalue of $\Lambda$ and $\sigma_i(J_T)$ the $i$th singular value of $J_T$. The ordered set $\lambda_1 \geq \lambda_2 \geq \cdots \geq \lambda_d$ is the Lyapunov spectrum. Positive exponents correspond to directions along which nearby trajectories diverge exponentially, and their number gives the dimension of the unstable manifold. The leading exponent sets the Lyapunov time $\lambda_1^{-1}$, which is the natural unit in which to report forecast horizons. The partial sums fix the Kaplan--Yorke dimension \cite{kaplan_chaotic_1979}
\begin{equation}
  D_{KY} = j + \frac{1}{|\lambda_{j+1}|}\sum_{i=1}^{j}\lambda_i,
  \qquad
  j = \max\Big\{ \ell : \textstyle\sum_{i=1}^{\ell}\lambda_i \geq 0 \Big\},
  \label{eq:kydim}
\end{equation}
which is conjectured to equal the information dimension of the attractor.

We compute the exponents with the standard QR-based algorithm \cite{benettin_lyapunov_1980a, benettin_lyapunov_1980b, eckmann_ergodic_1985, abarbanel_analysis_2012, pathak_using_2017}. An orthonormal set of perturbation vectors $Q_0$ is propagated by the local Jacobian at each step, $\tilde{Q}_{k+1} = \left.\partial F/\partial \zvec\right|_{\zvec_k} Q_k$, and re-orthonormalized every $\tau$ steps through the factorization $\tilde{Q} = Q R$, which keeps every perturbation vector from collapsing onto the leading direction. The exponents come out as time averages of the logarithms of the diagonal entries of $R$,
\begin{equation}
  \lambda_i = \lim_{K\to\infty} \frac{1}{K\tau\,\Delta t} \sum_{j=1}^{K} \ln \left(R_j\right)_{ii},
  \label{eq:qralg}
\end{equation}
where $R_j$ is the triangular factor from the $j$th re-orthonormalization.
For the ground-truth systems we obtain the Jacobian from the governing equations linearized about a reference trajectory on the attractor. For the learned dynamics we apply the algorithm to the closed-loop reservoir map \eqref{eq:Gmap} along a trajectory generated by the autonomous ESN. Differentiating \eqref{eq:Gmap} gives
\begin{equation}
  \frac{\partial G}{\partial \svec}\bigg|_{\svec_k}
  = (1-\alpha) I + \alpha\, \mathrm{diag}\!\left(1 - \tanh^2(\mathbf{a}_k)\right)\big(A + \sigma_{\mathrm{in}} W_{\mathrm{in}} W_{\mathrm{out}}\big),
  \label{eq:esnjac}
\end{equation}
where $\mathbf{a}_k = (A + \sigma_{\mathrm{in}} W_{\mathrm{in}} W_{\mathrm{out}})\svec_k + \mathbf{b}$ is the reservoir preactivation, so no numerical differentiation is needed. The reservoir map acts on $\Rb^{N}$ with $N$ much larger than the latent dimension $r$, and therefore has $N$ Lyapunov exponents. Following \cite{pathak_using_2017}, we compare the leading exponents of this map with those of the true system.

Replicating a spectrum is a more stringent test than forecasting accurately over a short horizon \cite{pathak_using_2017}. A model can track a particular trajectory well for several Lyapunov times while having an attractor of entirely different geometry, and even a perfect model of the dynamics separates from any reference trajectory under perturbation of the initial condition. Agreement between spectra therefore reflects the invariant structure of the learned dynamics rather than about the accuracy of any single forecast. Because Lyapunov exponents are invariant under smooth invertible changes of coordinates, agreement between the true and learned spectra is consistent with $\Enc$ recovering coordinates related to the true state space by a diffeomorphism. Agreement is necessary for such a relation but does not by itself establish it.

%=====================================================================
\section{Experimental systems}
\label{sec:systems}

\subsection{Kuramoto--Sivashinsky equation}
The Kuramoto--Sivashinsky (KS) equation is a canonical model of spatiotemporal chaos \cite{Sivashinsky1977, Kuramoto1978}. It governs the evolution of a scalar field $u(t, x)$ on a periodic domain of length $L$ and is given by
\begin{equation}
    u_t + u u_x + u_{xx} + u_{xxxx} = 0, \quad x \in [0, L], \quad u(x, 0) = u_0(x),
\end{equation}
with periodic boundary conditions $u(t, x+L) = u(t, x)$. The second-order term $u_{xx}$ is destabilizing and drives energy production at long wavelengths, while the fourth-order term $u_{xxxx}$ is dissipative and damps short wavelengths. The nonlinear advection $u u_x$ transfers energy between scales. The balance among these three effects produces chaotic dynamics whose complexity is set by the domain length $L$. As $L$ grows, more unstable modes fit inside the domain, the attractor dimension increases, and the dynamics become more complex \cite{HymanNicolaenkoZaleski1986}.

We study the equation at two domain lengths, $L = 22$ and $L = 44$. The $L = 22$ case is a standard benchmark in both the dynamical systems and machine learning literatures \cite{cvitanovic_state_2010, linot_deep_2020, linot_datadriven_2022}, and produces low-dimensional chaotic dynamics with a small number of positive Lyapunov exponents. The $L = 44$ case has more positive exponents and roughly twice the Kaplan--Yorke dimension (Table~\ref{tab:config}), giving a higher-dimensional attractor and more complex spatiotemporal behavior. The pair makes for a controlled comparison, since the governing equation is unchanged while the intrinsic complexity of the dynamics increases substantially. This provides a good test for how the forecasting pipeline and the spectrum replication scale with attractor dimension.

At each domain length we collect $m$ snapshots of the spatially discretized state $u_k \in \Rb^{n}$ at uniform time intervals from a trajectory on the attractor, after discarding an initial transient. Sparse measurements come from selecting $p$ components of the state vector at random spatial locations, with the same locations used across all snapshots. In all KS experiments we take $p=25$ and $n=128$, following \cite{OZALP2026118600}. The SHRED model is trained on trajectories of these $p$ measurements, with latent dimension $r$ and lag length $L_{\mathrm{lag}}$ listed in Table~\ref{tab:config}.

\subsection{Kolmogorov flow}
Our second benchmark is Kolmogorov flow, a two-dimensional incompressible flow driven by a sinusoidal body force and governed by the Navier--Stokes equations
\begin{equation}
    \frac{\partial \mathbf{u}}{\partial t} + (\mathbf{u} \cdot \nabla) \mathbf{u} = -\nabla P + \frac{1}{Re} \nabla^2 \mathbf{u} + \mathbf{f}, \quad \nabla \cdot \mathbf{u} = 0,
\end{equation} on a periodic domain $[0, 2\pi]^2$, where $\mathbf{u} = (u, v)$ is the velocity field, $P$ the pressure, $Re$ the Reynolds number, and $\mathbf{f}$ a steady forcing of the form $\mathbf{f} = (\sin(k_f y), 0)$ with forcing wavenumber $k_f$. We take $k_f = 4$ and study the flow at $Re = 30$, where it is quasi-periodic \cite{racca_predicting_2023, OZALP2026118600}. Data was obtained using the pseudo-spectral method. We take as the state the scalar vorticity $\omega = \partial_x v - \partial_y u$ on a uniform $48 \times 48$ grid, so that $n = 2304$. Compared with the KS equation this is a harder test of the sensing problem, since the gap between sensor count and state dimension is much wider, but an easier test of the modeling problem, since the dynamics are not chaotic.

We collect training data by sampling 100{,}000 snapshots at uniform time intervals from a trajectory on the attractor, again after discarding an initial transient. As in the KS experiments, the sensors measure the state itself: $p = 144$ point values of the vorticity at random grid locations that stay fixed across all snapshots, covering $6.25\%$ of the grid.

Quasi-periodic dynamics evolve on an invariant torus, and the Lyapunov spectrum of such a flow has no positive exponents. The spectrum diagnostic of Section~\ref{sec:lyap} is therefore uninformative here. We instead assess long-time fidelity geometrically, by comparing the manifold traced out by the autonomous ESN in latent space with the manifold traced out by the encoded latent trajectory.

%=====================================================================
\section{Results}
\label{sec:results}

The results build toward the central question of whether the learned latent dynamics replicate those of the underlying system. Section~\ref{sec:recon} summarizes reconstruction accuracy, comparing SHRED with a POD projection and a convolutional autoencoder (CAE), both of which see the full state. Section~\ref{sec:forecasting} compares forecasts from the SHRED- and CAE-based pipelines and shows that reconstruction accuracy does not determine forecast accuracy. Section~\ref{sec:lyapresults} shows that the learned latent dynamics of the KS equation reproduce the Lyapunov spectrum of the true system, and Section~\ref{sec:manifold} examines the invariant manifold of the learned dynamics for Kolmogorov flow.

\subsection{Reconstruction from sparse sensors}
\label{sec:recon}

We first establish how much of the field SHRED recovers from sparse sensors, relative to two references that see the full state and use the same latent dimension $r$. The first is a proper orthogonal decomposition (POD) projection onto the leading $r$ modes. POD gives the optimal rank-$r$ linear approximation of the data in the $\ell_2$ norm, so its error is a lower bound on the error of any linear reconstruction at that rank, including gappy POD from the same sensors. The second is a convolutional autoencoder (CAE) built along the lines of \cite{racca_predicting_2023, OZALP2026118600}, which serves as a full-state nonlinear reference. At fixed $r$, the gap between SHRED and the CAE measures the cost of observing the system only through sparse sensors. Table~\ref{tab:config} summarizes the three configurations and their reconstruction errors at the latent dimensions used in the rest of the paper, and Appendix~\ref{app:recon} shows how the errors depend on $r$.

\begin{table}[!ht]
    \centering
    \begin{tabular}{lccc}
        \toprule
        & KS, $L=22$ & KS, $L=44$ & Kolmogorov, $Re=30$ \\
        \midrule
        State dimension $n$ & $128$ & $128$ & $2304$ \\
        Sensors $p$ ($p/n$) & $25$ ($19.5\%$) & $25$ ($19.5\%$) & $144$ ($6.25\%$) \\
        Lags $L_{\mathrm{lag}}$ & 60 & 60 & 100 \\
        Latent dimension $r$ ($r/n$) & $8$ ($6.25\%$) & $18$ ($14.1\%$) & $7$ ($0.30\%$) \\
        Kaplan--Yorke dimension $D_{KY}$ & 5.235 & 9.919 & -- \\
        \midrule
        \multicolumn{4}{l}{Relative $\ell_2$ reconstruction error} \\
        \quad POD (full state) & 0.18 & 0.17 & 0.45 \\
        \quad SHRED (sensors) & 0.0047 & 0.021 & 0.021 \\
        \quad CAE (full state) & 0.0024 & 0.0037 & 0.011 \\
        \bottomrule
    \end{tabular}
     \caption{Configurations and reconstruction accuracy at the latent dimensions used throughout the paper. Percentages are relative to the state dimension $n$. $D_{KY}$ is the Kaplan--Yorke dimension \eqref{eq:kydim} of the true dynamics. Reconstruction errors are aggregate relative $\ell_2$ errors on test data. POD and the CAE take the full state as input, and SHRED takes only the sensor history.}
     \label{tab:config}
\end{table}

At these latent dimensions the latent state has $6.25\%$, $14.1\%$ and $0.30\%$ as many components as the discretized state for the KS equation at $L=22$ and $L=44$ and for Kolmogorov flow, respectively. These percentages depend on the grid resolution, so Table~\ref{tab:config} also lists the Kaplan--Yorke dimension of the true KS dynamics, a resolution-independent measure of the number of active degrees of freedom. For the KS equation the dimension of the inertial manifold on which the long-time dynamics evolve has also been estimated directly \cite{ding_estimating_2016, linot_deep_2020}.

In all three configurations SHRED reconstructs less accurately than the CAE, as one would expect given its far smaller input, but its error is much closer to the CAE's than to that of POD at the same latent dimension. SHRED working from sparse sensors therefore outperforms the best linear compression of the full state. The weakness of POD is expected here. Both systems are invariant under continuous spatial translations, along the periodic domain for the KS equation and in the streamwise direction $x$ for Kolmogorov flow, so their POD modes are close to Fourier modes in those directions, and coherent structures that drift, merge and split are poorly represented by any fixed low-dimensional linear subspace. The gap between SHRED and the CAE depends on the complexity of the dynamics. At $L=22$ the CAE's error is about half of SHRED's, while at $L=44$ it is about six times smaller. This widening is what one would expect, since the larger domain supports more active spatial modes and $25$ point measurements constrain the field correspondingly less.

\begin{figure}[!ht]
    \centering
    \subfloat[KS, $L=22$, $r=8$\label{fig:cae_rcae_recons_l22}]{%
        \includegraphics[width=0.49\linewidth]{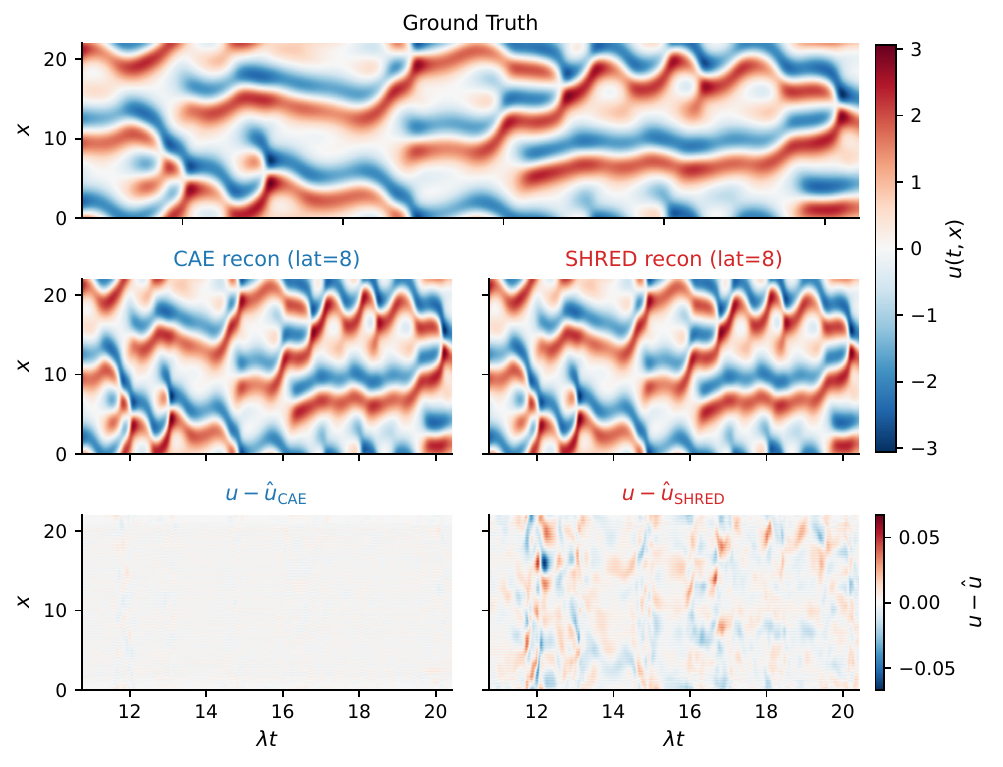}}
    \hfill
    \subfloat[KS, $L=44$, $r=18$\label{fig:cae_rcae_recons_l44}]{%
        \includegraphics[width=0.49\linewidth]{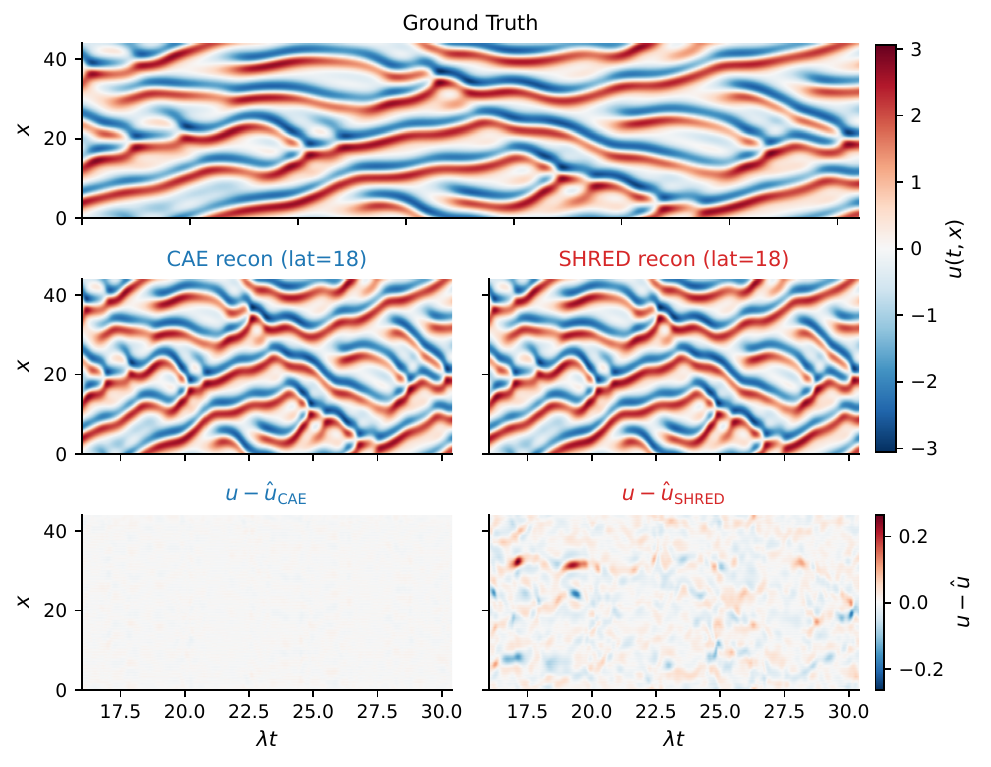}}
    \caption{Example reconstructions of the KS equation over the same interval of time, with horizontal axes in Lyapunov times, from the CAE, which takes the full state as input ($N_{in}=128$), and SHRED, which takes $25$ point sensors ($N_{in}=25$). In each panel the top row is the ground truth, the middle row shows the CAE (left) and SHRED (right) reconstructions, and the bottom row shows their pointwise errors. The SHRED error is larger but remains small compared with the field amplitude and shows no coherent structure.}
    \label{fig:cae_rcae_recons_ks}
\end{figure}

\begin{figure}[!ht]
    \centering
    \begin{overpic}[width=0.9\linewidth]{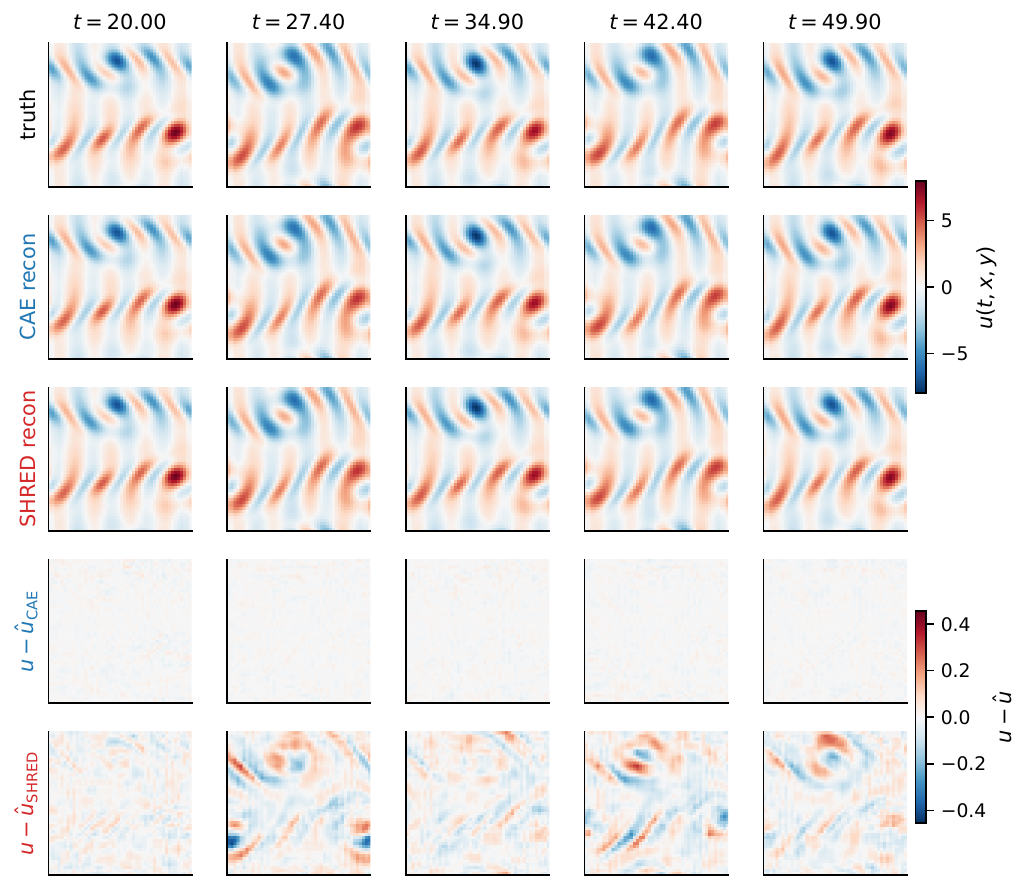}
        \put(-10, 24){\textcolor{blue}{{\scriptsize $N_{in} = 2304$}}}
        \put(-10, 7){\textcolor{red}{{\scriptsize $N_{in} = 144$}}}
    \end{overpic}
    \caption{Example reconstructions of the vorticity field of Kolmogorov flow at $Re=30$ over 50 nondimensional time units, at latent dimension $r=7$. Top: ground truth. Middle rows: reconstructions from the CAE, which takes the full vorticity field ($N_{in}=2304$), and from SHRED, which takes $144$ point sensors ($N_{in}=144$). Bottom rows: pointwise errors.}
    \label{fig:cae_rcae_recons_kol}
\end{figure}

Despite these differences, both approaches produce fields that are visually indistinguishable from the ground truth, as Figures~\ref{fig:cae_rcae_recons_ks} and~\ref{fig:cae_rcae_recons_kol} show. SHRED's pointwise error, while larger than the CAE's, remains small relative to the amplitude of the field and is not concentrated in any particular region or coherent structure. The same holds for Kolmogorov flow, even though its sensors cover only about $6\%$ of the grid, roughly a third of the relative sensor density of the KS experiments. SHRED therefore recovers the relevant spatiotemporal structure from sparse sensors alone. The remainder of the paper investigates whether its latent trajectory also reproduces the dynamics.

\subsection{Forecasting in the latent space}
\label{sec:forecasting}

\begin{figure}[!ht]
    \centering
    \begin{overpic}[width=0.9\linewidth]
        {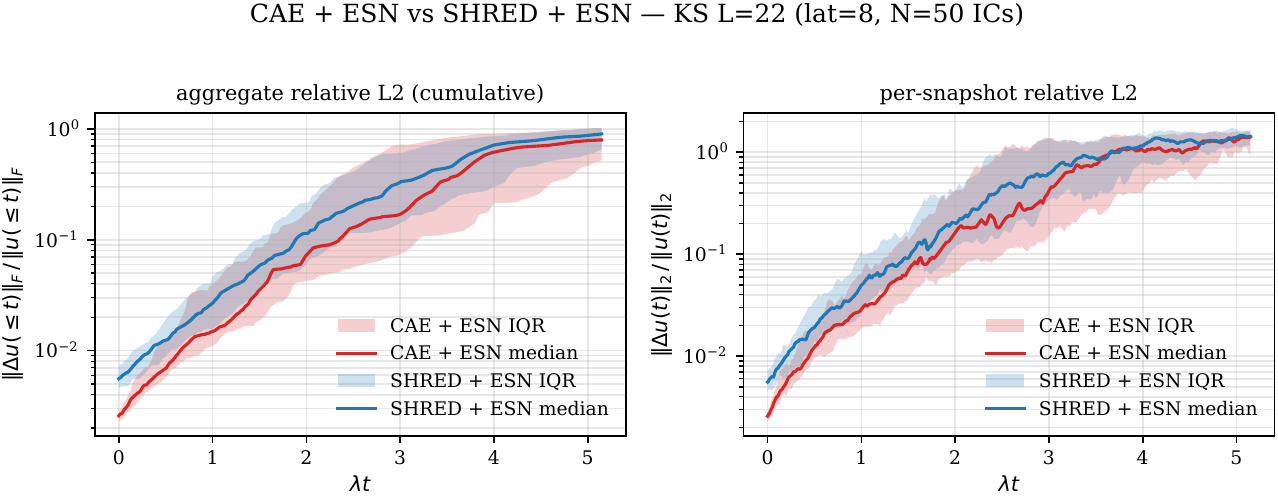}
        \put(35, 34){{\scriptsize SHRED $N_{in}$ = 25, CAE $N_{in} = 128$}}
        \put(18.911,36.610){\color{white}\rule{61.893\unitlength}{2.528\unitlength}}
        \put(20.396,36.610){\includegraphics[trim=118.973 224.507 308.656 0,clip,width=30.267\unitlength]{figures/ks_l22_panels.pdf}}
        \put(51.548,36.610){\includegraphics[trim=322.217 224.507 120.712 0,clip,width=27.772\unitlength]{figures/ks_l22_panels.pdf}}
        \put(50.831,37.514){\circle*{0.254}}
        \put(50.831,38.137){\circle*{0.254}}
    \end{overpic}
    \caption{Forecast error of the CAE + ESN (red) and SHRED + ESN (blue) pipelines for the KS equation at $L=22$ with $r=8$, over $N=50$ initial conditions drawn from a held-out test trajectory. From each initial condition the ESN runs autonomously and the decoder maps the latent forecast to the full field. The SHRED pipeline is initialized from $25$ sensor measurements and the CAE pipeline from the full state ($N_{in}=128$). Time is measured in Lyapunov times $\lambda_1 t$, with $\lambda_1 = 0.0504$. Left: cumulative relative $\ell_2$ error, i.e.\ the error norm accumulated over $[0,t]$ divided by the norm of the true field over the same interval. Right: relative $\ell_2$ error of each snapshot. Lines show medians and shaded bands interquartile ranges; lower curves indicate better forecasts. Since the KS field has zero spatial mean, a per-snapshot error near $\sqrt{2}$ means that the forecast and true states are effectively uncorrelated. Every forecast of a chaotic system eventually reaches this level, so the informative quantity is how long each pipeline takes to approach it. Here CAE + ESN has the lower error over roughly the first three Lyapunov times, consistent with its more accurate reconstructions, after which the two pipelines perform comparably.}
    \label{fig:forecast_22}
\end{figure}

\begin{figure}[!ht]
    \centering
        \begin{overpic}[width=0.9\linewidth]
        {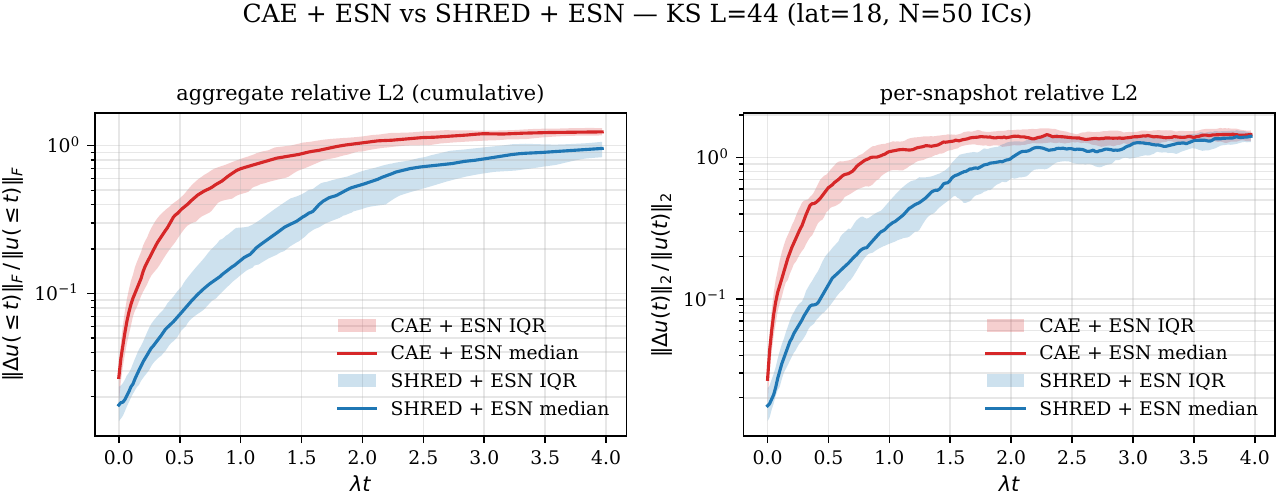}
        \put(35, 34){{\scriptsize SHRED $N_{in}$ = 25, CAE $N_{in} = 128$}}
        \put(18.321,36.595){\color{white}\rule{63.113\unitlength}{2.527\unitlength}}
        \put(19.805,36.595){\includegraphics[trim=115.400 224.507 312.476 0,clip,width=30.255\unitlength]{figures/ks_l44_panels.pdf}}
        \put(50.945,36.595){\includegraphics[trim=318.644 224.507 116.900 0,clip,width=29.005\unitlength]{figures/ks_l44_panels.pdf}}
        \put(50.228,37.499){\circle*{0.254}}
        \put(50.228,38.121){\circle*{0.254}}
    \end{overpic}
    \caption{Forecast error for the KS equation at $L=44$ with $r=18$, over $N=50$ initial conditions. The setup, metrics and colors are as in Figure~\ref{fig:forecast_22}, with $\lambda_1 = 0.0814$. Here SHRED + ESN (blue) has the lower median error over the whole horizon and approaches the uncorrelated level of about $\sqrt{2}$ more slowly than CAE + ESN (red), even though the CAE reconstructs about six times more accurately at this latent dimension (Table~\ref{tab:config}). Both pipelines lose accuracy faster than at $L=22$, as expected for the higher-dimensional attractor.}
    \label{fig:forecast_44}
\end{figure}

\begin{figure}[!ht]
    \centering
    \begin{overpic}[width=0.9\linewidth]
        {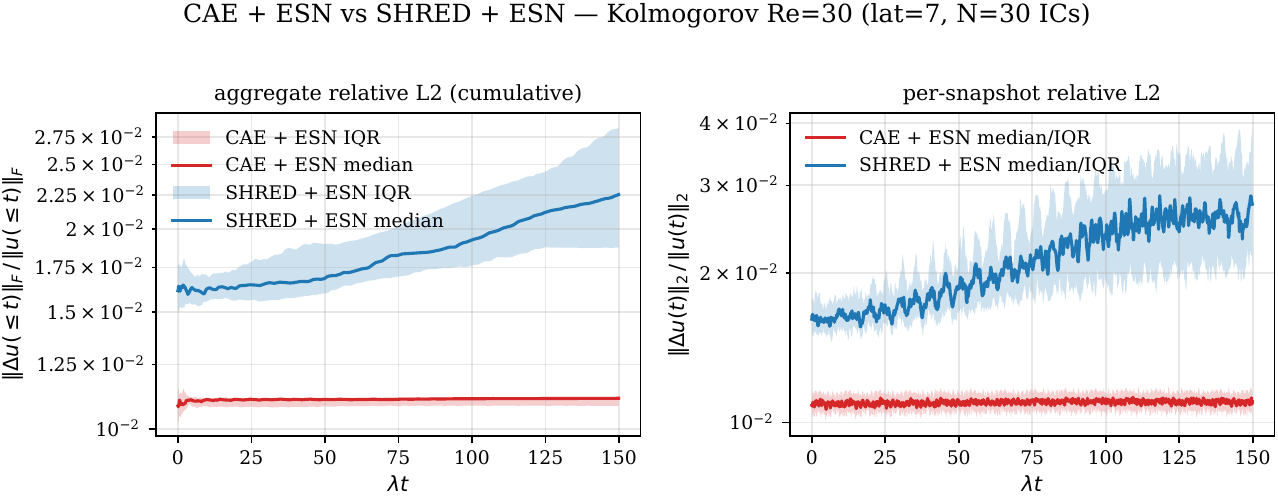}
        \put(35, 34){{\scriptsize SHRED $N_{in}$ = 144, CAE $N_{in} = 2304$}}
    \end{overpic}
    \caption{Forecast error of the CAE + ESN (red) and SHRED + ESN (blue) pipelines for Kolmogorov flow at $Re=30$ with $r=7$, over $N=30$ initial conditions; the metrics are as in Figure~\ref{fig:forecast_22}. The SHRED pipeline is initialized from $144$ sensor measurements and the CAE pipeline from the full vorticity field ($N_{in}=2304$). The horizontal axis is nondimensional time $t$, since the quasi-periodic flow has no positive Lyapunov exponent with which to rescale it. Note the vertical scales: all errors remain at the level of a few percent. Neither error grows exponentially, as expected for quasi-periodic dynamics. The CAE + ESN error stays essentially constant, while the SHRED + ESN error grows slowly, from about $1.6\%$ to $2.7\%$ per snapshot over $150$ time units.}
    \label{fig:forecast_re30}
\end{figure}

We turn next to the forecasting performance of the CAE + ESN and SHRED + ESN pipelines. In both cases, an echo state network propagates the latent state forward on its own and the decoder recovers the full state. The two pipelines are identical downstream of the latent representation and differ only in how that representation is produced.

Figure \ref{fig:forecast_22} shows the forecasting error for the KS system at $L=22$, evaluated over $N=50$ randomly selected initial conditions from a withheld test trajectory. CAE + ESN is more accurate over roughly the first three Lyapunov times, as its better reconstruction would suggest, after which the two approaches forecast about equally well. Figure \ref{fig:forecast_44} shows the same comparison at $L=44$, where the ordering flips. SHRED + ESN now outperforms CAE + ESN, and the advantage holds across the full forecast horizon of four Lyapunov times. Valid prediction times are shorter for both approaches than at $L=22$, as one would expect given the larger attractor dimension.

The reversal at $L=44$ is the more interesting of the two results, because $L=44$ is where the CAE's reconstruction advantage was largest, about a factor of six. An architecture that reconstructs six times worse nonetheless forecasts better, indicating that reconstruction accuracy and forecastability are distinct properties of a latent representation and the first does not imply the second. The CAE encoder is applied snapshot by snapshot, so its latent coordinates are optimized for reconstruction with nothing tying consecutive latent states together, and the resulting latent trajectory can be irregular in time even when every individual reconstruction is excellent. The GRU encoder computes each latent state from a long window of sensor history that overlaps almost entirely with the window for the previous state, and at deployment it runs as the recurrence \eqref{eq:gru} driven by the sensor stream. Its output is therefore a smoothly varying trajectory to begin with, and the ESN's job of learning an autonomous map on that trajectory is better posed as a result. The recurrent encoder gives up reconstruction fidelity in exchange for
temporal regularity, and in the considered regime that appears to result in better forecasting performance.

Figure \ref{fig:forecast_re30} gives the forecasting comparison for Kolmogorov flow at $Re = 30$ over $N=30$ initial conditions. Here CAE + ESN has the lower median error throughout the forecast horizon. Both approaches still forecast accurately over long horizons of 150 nondimensional time units, with errors that remain at the level of a few percent and grow slowly, if at all, rather than exponentially as they do for KS. That is consistent with the flow being quasi-periodic at this Reynolds number. Propagating a trajectory on a torus demands little of the temporal structure of the latent representation, so this case does not test the mechanism behind the $L=44$ result. We include it to show that the pipeline extends to two-dimensional fields and to a much sparser sensing regime, and we rest the claim about the benefit of recurrence on the KS results.

Taken together, the three configurations show a consistent ordering. SHRED + ESN trails CAE + ESN on the quasi-periodic flow, matches it on the low-dimensional chaotic attractor at $L=22$, and overtakes it on the higher-dimensional attractor at $L=44$. The relative standing of the recurrent pipeline thus improves as the dynamics become more complex. Three configurations are too few to establish a trend, but the ordering is what one would expect if reconstruction accuracy dominates forecast error when the dynamics are simple and the temporal structure of the latent representation dominates when they are not.

\subsection{Replication of the Lyapunov spectrum}
\label{sec:lyapresults}

\begin{figure}[!ht]
    \centering
    \includegraphics[width=0.9\linewidth]{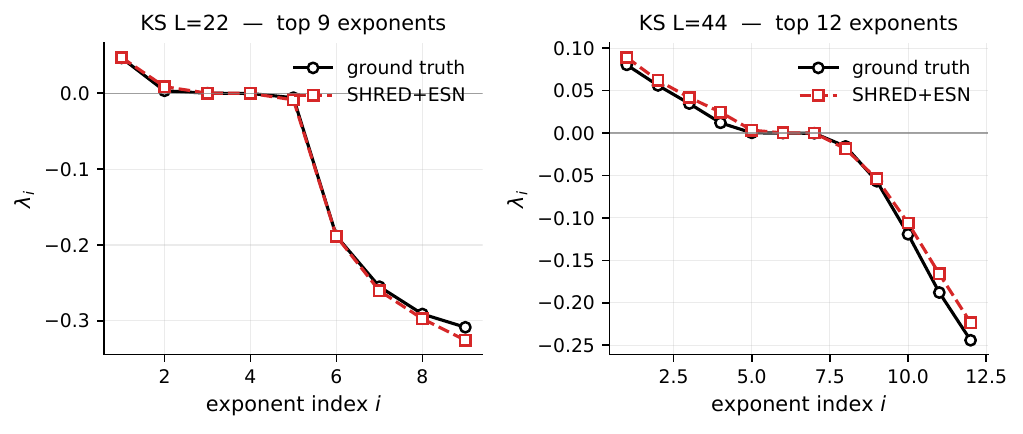}
    \caption{Lyapunov spectra of the true KS dynamics (black) and the learned SHRED + ESN latent dynamics (red, dashed) for $L=22$ (left, top 9 exponents) and $L=44$ (right, top 12 exponents). The learned spectra are augmented with an additional zero exponent; the true spectra are not (see Section~\ref{sec:lyapresults}).}
    \label{fig:lyap}
\end{figure}

The forecasting results show that the latent representation supports accurate prediction over finite horizons. We now turn to the central question of the paper, whether the learned latent dynamics replicate the dynamics of the true system. Finite-horizon accuracy does not answer this question, since it is compatible with a learned model whose attractor differs considerably from the true one. We therefore compute the Lyapunov spectrum, which characterizes the invariant structure of the dynamics rather than the accuracy of any single trajectory.

Figure \ref{fig:lyap} gives the Lyapunov spectra for the KS equation at both domain lengths, computed from the true dynamics and from the learned ESN by the procedure of Section~\ref{sec:lyap}. The learned spectra, but not the true ones, carry an additional zero exponent, following Pathak et al.\ \cite{pathak_using_2017}. At both $L=22$ and $L=44$ the SHRED + ESN spectrum matches the ground truth closely across the whole range of exponents considered, reproducing the positive exponents that determine the valid prediction time along with the near-zero and negative parts of the spectrum. The corresponding Kaplan--Yorke dimensions are $D_{KY} = 5.235$ (true) and $5.252$ (learned) at $L=22$, and $9.919$ and $10.250$ at $L=44$. We omit spectra for Kolmogorov flow, since the quasi-periodic dynamics at $Re=30$ have no positive exponents and the comparison would say little. Section~\ref{sec:manifold} examines the invariant structure of that case geometrically instead.

The Lyapunov spectrum is invariant under smooth invertible changes of coordinates, so two systems related by a diffeomorphism share a spectrum, and a learned model whose latent coordinates were nothing more than a convenient compression would have no particular reason to reproduce one. The autonomous latent time stepper recovers the spectrum of the full high-dimensional system while the encoder supplying its latent states receives only $25$ of the $128$ state components as input. We read this as evidence that the recurrent encoder has found coordinates related to the true attractor by a diffeomorphism, which is what the delay-embedding argument of Section~\ref{sec:intro} suggests. Spectrum agreement is a necessary consequence of such a relation rather than a proof of it, but it is a demanding one, since it constrains the rates of expansion and contraction of the learned dynamics along every direction compared. It also extends the result of Pathak et al.\ \cite{pathak_using_2017}, who demonstrated spectrum replication for a reservoir driven by the full KS state, to the case where the reservoir sees only a latent state distilled from sparse point measurements. The dynamical content of the field survives both the sparse sampling and the nonlinear compression.

\subsection{Latent manifold of Kolmogorov flow}
\label{sec:manifold}

\begin{figure}[!ht]
    \centering
    \includegraphics[width=\linewidth, trim={0 0 0 34}, clip]{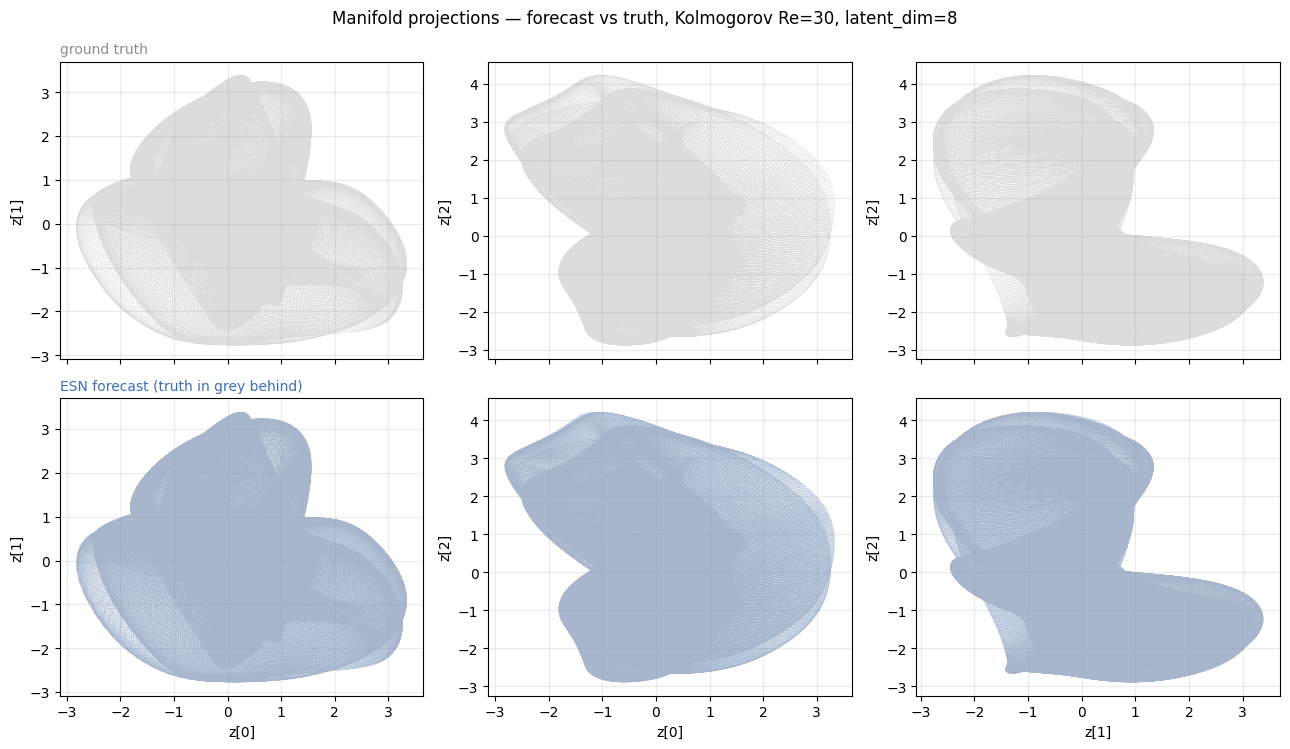}
    \caption{Two-dimensional projections of the SHRED latent manifold for Kolmogorov flow at $Re=30$. Axes are the first three standardized latent coordinates, indexed from zero. Top row: latent trajectory produced by the SHRED encoder from sensor measurements. Bottom row: autonomous SHRED + ESN latent forecast (blue), drawn over the encoded trajectory (grey).}
    \label{fig:manifold_kol}
\end{figure}

For quasi-periodic dynamics the natural test of long-time fidelity is whether the learned model settles onto the same invariant set as the true system. A time stepper can keep the error small over a finite horizon, as in Figure~\ref{fig:forecast_re30}, and still fail in this respect. It can relax onto a periodic orbit, which carries fewer incommensurate frequencies than the torus, or it can wander slowly off the invariant set, and neither failure need show up in the relative $\ell_2$ error over $150$ time units. We therefore look directly at the latent geometry.

Figure~\ref{fig:manifold_kol} shows projections of the latent trajectory onto pairs of its first three coordinates. The top row gives the trajectory obtained by encoding sensor measurements with the SHRED encoder. The trajectory densely fills a bounded region with sharp outer boundaries and several folds, and it winds around the region in closely spaced bands rather than scattering across it. This is the appearance one expects from a trajectory on a torus that has been embedded smoothly in $\Rb^{r}$ and then projected onto a plane. The encoder was trained only to reconstruct the flow field through \eqref{eq:shredloss}, so this organization was not imposed. As with the spectrum agreement for KS, it comes from the recurrent parameterization acting on a delay vector of measurements.

The bottom row overlays the autonomous ESN forecast on the encoded trajectory. In every projection the forecast covers the same region, follows the same outer boundaries, and reproduces the folds and lobes of the encoded manifold, including the narrow tail near $z[1] \approx -1$, $z[2] \approx -2.5$ in the right-hand panel. The forecast does not collapse onto a closed curve, which would indicate that the ESN had learned a periodic approximation of the dynamics, and it does not spread beyond the manifold, which would indicate drift off the invariant set. The ESN has been trained only on one-step prediction of the latent state and is iterated here without any sensor input, so agreement at this level shows that it has learned the invariant set of the latent dynamics and not merely a good short-horizon predictor.

Two caveats apply. The comparison is qualitative, and a projection onto three of $r$ coordinates can hide discrepancies in the remaining directions. A quantitative comparison, for example of the frequency content of the latent trajectory or of distances between the invariant measures, would sharpen the result. Even so, taken with the small and slowly varying forecast error of Figure~\ref{fig:forecast_re30}, the figure indicates that the SHRED + ESN pipeline reproduces the long-time behavior of the flow from sensors covering only about $6\%$ of the grid.

%=====================================================================
\section{Discussion}
\label{sec:discussion}

We have shown that latent dynamics learned from a recurrent encoding of sparse sensor measurements replicate the invariant dynamical structure of the underlying system, not merely its short-term trajectory. On the Kuramoto--Sivashinsky equation, the autonomous latent time stepper recovers the leading Lyapunov exponents of the true dynamics at both $L=22$ and $L=44$, working from $25$ point sensors on a $128$-point grid and a latent state with at most $14\%$ as many components as the full state. On Kolmogorov flow at $Re=30$ the pipeline extends to a two-dimensional field and a much sparser sensing regime, and the autonomous time stepper stays on the smooth manifold traced out by the encoded latent trajectory and reproduces its geometry over long horizons.

The fact that a network trained to reconstruct a field from sparse measurements learns a latent code sufficient for reconstruction has been demonstrated before \cite{williams_data-driven_2022, ebers2023leveraging}. That the same code should also serve as a faithful coordinate system for the dynamics is less obvious, since nothing in the training objective \eqref{eq:shredloss} specifically enforces it. The spectrum agreement suggests that it does, and the delay-embedding argument offers an explanation: the recurrent encoder consumes the very delay vector Takens' theorem identifies as diffeomorphic to the system state, so the representation it learns can inherit the invariants of the original attractor. In practical terms, system identification from sparse sensors need not pass through full-state reconstruction as an intermediate step. The dynamics can be characterized directly, in a coordinate system the sensors themselves induce.

Another result of our work concerns encoding strategies. An autoencoder-based pipeline takes the full state as encoder input. Thus, a sensor-constrained deployment has to reconstruct the field before the forecasting pipeline can run at all. That inverse problem is not trivial, and errors it introduces propagate through the encoder into the latent representation and can corrupt a downstream forecast. The recurrent formulation sidesteps this by construction, since the encoder works directly on sensor trajectories and its latent state is conditioned on the available measurements from the start. There is also a subtler benefit, visible in the forecasting comparison of Section~\ref{sec:forecasting}. Because the recurrent layer is itself a model of temporal dynamics, the latent trajectory it produces is better conditioned for autonomous propagation than one assembled from independently encoded snapshots. At $L=44$ that effect is large enough to overturn a sixfold reconstruction advantage in the CAE's favor. Reconstruction error is therefore a poor criterion on its own for choosing a reduced-order model meant for forecasting, and the temporal structure of the latent representation is also of critical importance.

Several limitations offer further directions for exploration. The KS equation was studied at two domain lengths, and the drop in forecasting accuracy from $L=22$ to $L=44$ raises the question of how the approach scales to systems with much higher-dimensional attractors. These experiments do not establish what sensor budget is needed to sustain spectrum replication as attractor dimension grows. The Kolmogorov flow was studied at a single Reynolds number where the flow is quasi-periodic, so it tests primarily the sensing and reconstruction parts of the pipeline, and our evidence that the learned dynamics are faithful there rests on a qualitative comparison of latent manifolds rather than on a dynamical invariant. Higher Reynolds numbers, where the attractor dimension grows and the flow becomes turbulent, are the natural next test. More broadly, the systems here are statistically stationary, and we have not explored behavior under intermittency, transient chaos, or slowly drifting parameters. The construction also assumes sensors at fixed spatial locations sampled on a fixed temporal grid, since the selection operator $C$ and the sampling interval are both baked into the learned encoder. Mobile sensors have been handled for reconstruction \cite{ebers2023leveraging, tomasetto_reduced_2025}, but whether dynamics replication survives with mobile or asynchronous measurements remains open. Relatedly, the decoder is tied to the grid on which it was trained. Replacing it with a coordinate-based decoder, in the manner of the trunk network of DeepONet \cite{lu_learning_2021}, would allow the field to be queried at arbitrary locations, and pairing such a decoder with an encoder that accepts variable sensor layouts would move the approach toward operator learning in the sense of \cite{kovachki_neural_2023}. Finally, the ESN's hyperparameter sensitivity was manageable for the systems studied here but may matter more as the latent dynamics become more complex, which motivates hybrid reservoir approaches that fold in partial physical knowledge \cite{pathak_hybrid_2018, wikner_combining_2020} and regularization strategies that stabilize autonomous reservoir predictions \cite{wikner_stabilizing_2024}.

\section*{Acknowledgments} 

The authors acknowledge support from the National Science Foundation AI Institute in Dynamic Systems (grant number 2112085). JNK acknowledges support from the Air Force Office of Scientific Research (FA9550-24-1-0141). 

%=====================================================================
\section*{Declarations}

\noindent\textbf{Conflict of interest.} The authors declare no competing
interests.

%=====================================================================
\appendix
\renewcommand{\thefigure}{A\arabic{figure}}
\setcounter{figure}{0}

\section{Reconstruction error as a function of latent dimension}
\label{app:recon}

\begin{figure}[!ht]
    \centering
    \begin{overpic}[width=0.9\linewidth]{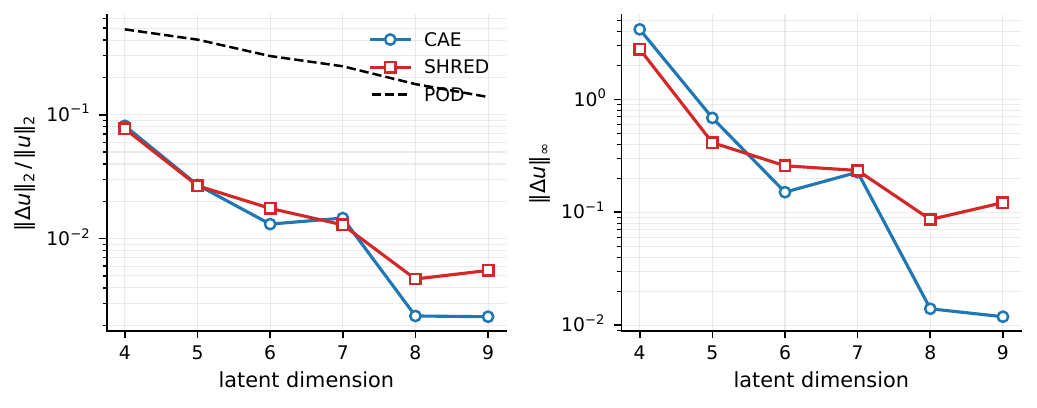}
        \put(25,40){SHRED $N_{in}$ = 25, CAE and POD $N_{in} = 128$}
    \end{overpic}
    \caption{Reconstruction error as a function of latent dimension $r$ for the KS equation at $L=22$, for POD and the CAE, which take the full state as input ($N_{in}=128$), and SHRED, which takes $25$ point sensors ($N_{in}=25$). Left: aggregate relative $\ell_2$ error. Right: maximum pointwise error.}
    \label{fig:cae_rcae_errors_l22}
\end{figure}

\begin{figure}[!ht]
    \centering
    \begin{overpic}[width=0.9\linewidth]{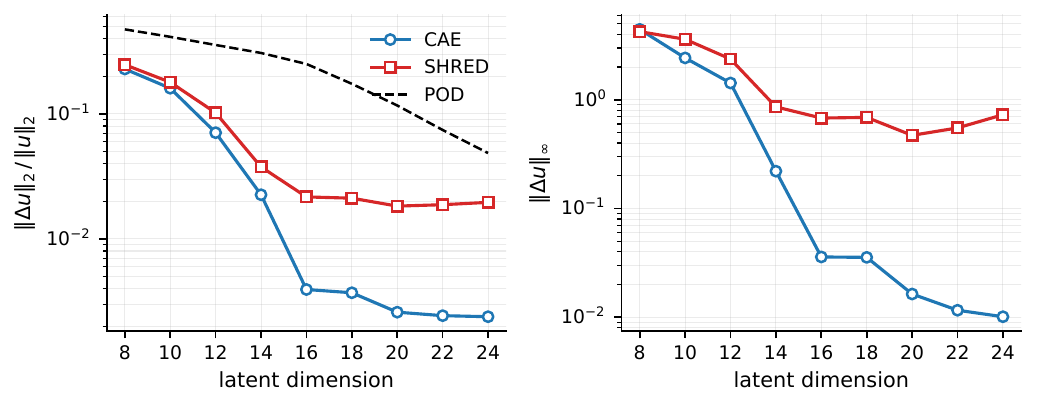}
        \put(25,40){SHRED $N_{in}$ = 25, CAE and POD $N_{in} = 128$}
    \end{overpic}
    \caption{Reconstruction error as a function of latent dimension $r$ for the KS equation at $L=44$. Inputs and panels are as in Figure~\ref{fig:cae_rcae_errors_l22}.}
    \label{fig:cae_rcae_errors_l44}
\end{figure}

\begin{figure}[!ht]
    \centering
    \begin{overpic}[width=0.9\linewidth]{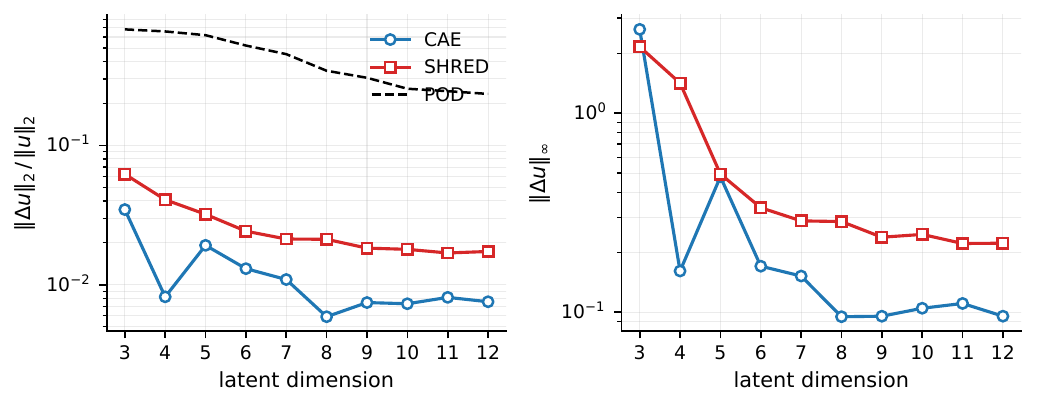}
        \put(25,40){SHRED $N_{in}$ = 144, CAE and POD $N_{in} = 2304$}
    \end{overpic}
    \caption{Reconstruction error as a function of latent dimension $r$ for the vorticity field of Kolmogorov flow at $Re=30$, for POD and the CAE, which take the full field as input ($N_{in}=2304$), and SHRED, which takes $144$ point sensors ($N_{in}=144$). Left: aggregate relative $\ell_2$ error. Right: maximum pointwise error.}
    \label{fig:cae_rcae_errors_kol}
\end{figure}

Figures~\ref{fig:cae_rcae_errors_l22}--\ref{fig:cae_rcae_errors_kol} show the reconstruction error of POD, the CAE and SHRED as a function of latent dimension $r$ for the three configurations. POD and the CAE take the full state as input, and SHRED takes only the sensor history. At $L=22$ the CAE and SHRED are comparable at low latent dimensions and both plateau beyond $r=8$, with the CAE settling at roughly half of SHRED's aggregate relative $\ell_2$ error. At $L=44$ the CAE is consistently better for $r>14$, by a factor of about five to eight. For Kolmogorov flow the CAE is again more accurate across latent dimensions, and both approaches plateau around $r=7$. In each case the POD error decreases only slowly with $r$, and SHRED's error lies much closer to the CAE's than to that of POD over the range of latent dimensions shown.

\bibliographystyle{unsrt}
\bibliography{shredesnbib}
\end{document}